\documentclass[manuscript,screen]{acmart} 

\usepackage{booktabs}
\usepackage{tabularx}

\usepackage{hyperref}
\usepackage{xcolor}
\usepackage{todonotes}

\AtBeginDocument{%
  }

\setcopyright{acmlicensed}
\copyrightyear{2026}
\acmYear{2018}
\acmDOI{XXXXXXX.XXXXXXX}

\acmConference[CS6206]{Human Computer Interaction}{Semester 2,
  2026}{Singapore}
\acmBooktitle{CS6206: Human Computer Interaction, AY25/26 Semester 2, National University of Singapore}
\acmISBN{978-1-4503-XXXX-X/18/06}

\begin{document}

\title{AI-Driven Urge Regulation Assistant (AURA): Designing Preemptive Smart Wearables for Smoking Cessation}


\author{Timothy Antoni}
\affiliation{%
  \institution{National University of Singapore}
  \streetaddress{13 Computing Drive}
  \country{Singapore}
}
\email{timothy.antoni@u.nus.edu}

\author{Kevin F. Y. Lam}
\affiliation{%
  \institution{National University of Singapore}
  \streetaddress{13 Computing Drive}
  \country{Singapore}
}
\email{lamfy@u.nus.edu}

\author{Junyi Hou}
\affiliation{%
  \institution{National University of Singapore}
  \streetaddress{13 Computing Drive}
  \country{Singapore}
}
\email{hhh@u.nus.edu}

\author{Wei Soon Cheong}
\affiliation{%
  \institution{National University of Singapore}
  \streetaddress{13 Computing Drive}
  \country{Singapore}
}
\email{weisoon.cheong@u.nus.edu}

\renewcommand{\shortauthors}{Antoni et al.}

\begin{abstract}
Smoking cessation remains a complex self-regulatory challenge, often undermined by cue-triggered cravings arising from habitual, emotional, and social contexts. Existing wearable cessation systems are largely reactive, detecting smoking events only after lapses occur and thus missing the critical window for timely intervention. Guided by cue-reactivity theory and Self-Determination Theory (SDT), we argue that future digital health systems should shift from retrospective feedback toward proactive craving management that supports user autonomy and competence.
To inform the design of such systems, we present a formative pilot qualitative study involving semi-structured interviews with 10 smokers and ex-smokers from a multi-ethnic Asian population, a demographic underrepresented in current wearable AI and smoking cessation research. Through an iterative co-design interview process, we examine (1) the barriers and facilitators to smoking cessation and (2) the desired design features of wearable systems targeting cravings before lapses occur.
Our findings identify key craving predictors across habitual, emotional, and social dimensions, alongside user-preferred intervention strategies such as social accountability, personalized support, rewards, and just-in-time distraction techniques. Participants also emphasized the importance of trust, privacy, personalization, and non-judgmental interaction styles.
These insights directly inform the design of AURA, a novel smartwatch-smartphone ecosystem for proactive craving detection and intervention using commercially available wearable devices.
\end{abstract}




\maketitle

\section{Introduction}
Imagine you have 3 friends. 
\textbf{Narita} is exciting at first, she checks in occasionally for a few months, then disappears. 
\textbf{Curran} stays in your life long-term, but only points out your mistakes after they happen. 
\textbf{Aurora}, in contrast, understands your habits and warns you before you make a poor decision. 
Among these, Aurora is the one most likely to create lasting positive change. 
This analogy reflects a fundamental challenge in the design of smoking cessation technologies: how systems interact with users over time, and whether they support action before or after a lapse occurs.

Smoking remains the primary cause of preventable death and disease~\cite{Horvath2024-zz, GlobalTobaccoControl, Papanicolas2025-re}, costing roughly \$1.4T annually~\cite{Goodchild2018-xw, Xu2015-ji}. 
Traditional smoking cessation methods such as Nicotine Replacement Therapy (NRT) and Cognitive Behavioral Therapy (CBT) suffer from low long-term abstinence retention and issues of impracticality~\cite{Young2010-xx, Etter2006-zk, Pekel2015-ar, Agboola2015-us, Mills2010-hj, Halperin2009-ba, Yasin2012-lk, Rigotti2010-tu, Baumeister1994-wi, haug2013efficacy}. 
From an HCI perspective, these limitations can be attributed to suboptimal interaction patterns.
Interventions are typically short-lived (3--6 months) and sparse in engagement (e.g. weekly sessions), resembling Narita, present initially but absent for long-term behaviour change.
This lack of sustained, context-aware support contributes to modest 52-week abstinence rates of around 20\%~\cite{Young2010-xx, Etter2006-zk, Pekel2015-ar, Agboola2015-us, Mills2010-hj, Halperin2009-ba, Yasin2012-lk, Rigotti2010-tu, Baumeister1994-wi, haug2013efficacy}.

Wearable AI systems offer a promising alternative by enabling continuous, passive, and long-term monitoring of users' physiological and behavioral states.
However, current systems largely resemble Curran: they detect smoking events after they occur, providing retrospective feedback.
While such systems promise strong potential for long-term monitoring, they fail to intervene during the critical window preceding a lapse. 
This limitation is particularly problematic when viewed through cue-reactivity theory~\cite{niaura1988relevance, betts2021meta}, which explains how spontaneous cravings triggered by contextual cues can rapidly override long-term intentions. 
Intervening only after a smoking event neglects the moment when support is most needed. 
Indeed, evidence demonstrates that early prevention leads to significantly better results in smoking reduction than systems designed to rebuke users retrospectively~\cite{Shadel2011-ct, Brown2005-lp, Collins2010-ue, Livingstone-Banks2019-kq}. 
To date, only AutoSense~\cite{Ertin2011-rh} attempted to prevent smoking in advance, but it utilizes a suite of six sensors that are impractical for long-term daily use, potentially reducing long-term cessation effectiveness; akin to traditional smoking cessation methods. 
A scalable solution must therefore rely on devices that users already possess, such as commercial smartwatches and smartphones.

Furthermore, existing wearable systems rarely personalize craving prediction to individual users, despite the availability of rich multi-modal data from commercial devices~\cite{El-Toukhy2025-yn, Dumortier2016-ro, Haasova2016-kg, Nadell2015-dl, Libby2012-ki, Ashare2012-oh, Erblich2011-rw}. 
Current approaches often treat users as part of a general population distribution, overlooking the highly individualized nature of craving triggers.
For example, one user's cravings may follow stressful meetings, while another's may occur during social contexts~\cite{Horvath2024-zz}. 
Without personalization, interventions risk being mistimed or irrelevant, thereby reducing their effectiveness.
From the perspective of Self-Determination Theory (SDT)~\cite{deci2012self, choi2014smoking}, this lack of personalization may also undermine users' sense of autonomy and competence, limiting their ability to meaningfully engage in behaviour change.

To address these gaps, we propose the AI-Driven Urge Regulation Assistant (AURA), a novel smartwatch-smartphone system aimed at accurately predicting craving tendencies, while simultaneously enabling personalized interventions for sustainable cessation practices. 
AURA embodies the role of Aurora: a system that understands the user and provides proactive, context-aware support to prevent lapses before they occur.

However, designing such a system requires a grounded understanding of the factors that drive smoking cravings, the intervention strategies that users find effective, and the practical considerations for wearable cessation tools.
To this end, we conduct a formative qualitative study with smokers from Asian populations, who remain underrepresented in existing smoking cessation research~\cite{Maleki2022-pl, drenkow2025detecting}---guided by the following research question: \textit{What are the key predictors of smoking cravings as experienced by smokers, what intervention strategies do they find effective, and what design considerations should guide the development of proactive wearable cessation systems?}

Through semi-structured interviews, we elicit smokers' lived experiences with craving triggers, their perspectives on cessation strategies, and their attitudes toward wearable health technologies.
This formative study situates itself as a necessary precursor within a broader three-stage development plan for AURA: (1) wearable data collection, (2) predictive model development, and (3) an end-to-end intervention study.
The qualitative insights obtained here will directly inform the design of AURA's sensing and intervention components in subsequent stages, while also contributing to the understanding of smoking behavior within underrepresented Asian demographics.

Our research questions are outlined below:
\begin{enumerate}
    \item RQ1: What are the barriers and facilitators to effective smoking cessation in a multi-ethnic Asian population?
    \item RQ2: What are the desired design features of wearable smoking cessation systems that specifically target craving management rather than post-smoking intervention?
\end{enumerate}

The contributions of this paper are summarized as follows:
\begin{enumerate}
    \item A qualitative characterization of smoking craving predictors across emotional, social, and contextual dimensions, grounded in smokers' lived experiences.
    \item User-derived intervention themes for smoking cessation, including distraction, accountability, achievement tracking, and voice of reason.
    \item Design considerations for proactive wearable cessation systems, addressing personalization, privacy, trust, and empowerment, with implications for the design of AURA.
\end{enumerate}

\section{Related Work}

\subsection{Drivers and Inhibitors of Smoking Cessation}
Understanding the drivers and inhibitors of smoking cessation is essential for designing effective behavioral and technological interventions. Prior work~\cite{luBarriersFacilitatorsSmoking2025, nguyenDesiresNeedsQuitting2024, leppinExploringPerspectivesDigital2025} has consistently shown that smoking cessation is influenced by a combination of individual, social, and contextual factors, including craving intensity, self-efficacy, stress, social support, habitual routines, and environmental triggers.

At the individual level, barriers commonly include low motivation, nicotine dependence, stress regulation needs, fear of withdrawal symptoms, and low confidence in one’s ability to quit. Self-efficacy, in particular, has been repeatedly identified as a key determinant of successful cessation outcomes. Individuals who perceive themselves as capable of resisting cravings are significantly more likely to sustain abstinence over time~\cite{kerrBreakingHabitQualitative2013}.

At the social level, facilitators often include supportive family members, romantic partners, peers, and healthcare professionals. Encouragement from close social circles not only strengthens motivation but also provides accountability and emotional reinforcement during high-risk craving episodes. Conversely, smoking-conducive environments---such as workplace smoke breaks, social drinking contexts, and peer smoking norms---can act as powerful inhibitors to cessation.

These findings are highly relevant to wearable intervention design, as they suggest that effective systems should go beyond craving detection alone and incorporate socially aware, context-sensitive, and motivationally adaptive support mechanisms.

\subsection{Psychological Theories Informing Smoking Cessation Design}
The design of our interview protocol and proposed system is grounded in established psychological theories of smoking behavior and behavior change.
First, cue-reactivity theory suggests that smoking-related environmental or internal cues can automatically evoke craving responses through learned associations formed over repeated smoking episodes. These cues may include stress, specific locations, social situations, or habitual daily routines. Once triggered, cravings can rapidly override long-term intentions to quit, increasing relapse risk.

Second, Self-Determination Theory (SDT) provides a strong framework for understanding sustained cessation behavior. SDT posits that long-term behavior change is more likely when interventions support the psychological needs of autonomy, competence, and relatedness. In the smoking cessation context, autonomous motivation—quitting because it aligns with personal values and goals—has been shown to predict more durable abstinence than externally imposed pressure~\cite{niemiecSelfdeterminationTheorySmoking2023}.

This theoretical perspective directly informs our system design philosophy: rather than delivering prescriptive alerts, the wearable should empower users with awareness, personalized choices, and supportive interventions that reinforce intrinsic motivation.

Additionally, elements of Social Cognitive Theory, particularly self-efficacy and outcome expectancy, inform our interview questions regarding trust, perceived usefulness, and confidence in managing cravings~\cite{kerrBreakingHabitQualitative2013}.

A comprehensive overview of all the frameworks used in constructing the interview questions are in Appendix 1.

\subsection{Empowering Users Through Preemptive Craving-State Awareness}
Smoking relapse is often driven by cue-triggered cravings that emerge from environmental or internal stimuli associated with past smoking behavior~\cite{Yasin2012-lk, Brown2005-lp, Collins2010-ue, Erblich2011-rw}. According to cue-reactivity theory, these cues can automatically evoke strong craving responses that frequently override an individual’s long-term goal of quitting, particularly when the craving arises outside of conscious awareness.
Increasing awareness of these emerging craving states can therefore play a critical role in interrupting the cue–craving–smoking cycle before a lapse occurs.

In the context of wearable technologies, continuous sensing of physiological and contextual signals creates opportunities for systems to infer and communicate such craving states in real time. From the perspective of SDT, providing users with insight into their own craving patterns can support psychological needs for autonomy and competence, enabling informed self-regulation rather than reliance on externally imposed reminders.

Consequently, wearable AI systems that surface craving awareness represent a promising HCI approach for empowering users to actively regulate their behavior and sustain long-term cessation.
Existing assistive technologies for smoking cessation predominantly focus on post-event detection and intervention, which may limit their ability to address the underlying craving processes that drive relapse~\cite{Shadel2011-ct, Brown2005-lp, Collins2010-ue, Livingstone-Banks2019-kq}. Systems such as AutoSense~\cite{Ertin2011-rh, hernandezSmokingCessationUsing2021, vinciProximalEffectsJustinTime2025} have been used in just-in-time intervention studies, but their reliance on specialized sensing hardware limits real-world scalability.

To date, limited work has explored the technical feasibility of preemptive craving-state detection using solely wearable data, and even fewer studies have examined the long-term behavioral implications of such systems in naturalistic longitudinal settings.
Our work addresses this gap by focusing specifically on user-centered design requirements and perceived acceptability of preemptive craving interventions.

\subsection{Passive and Long-Term Monitoring Through Wearable Health Technologies}
The integration of digital technologies into smoking cessation has long been anticipated as a promising direction for future healthcare delivery~\cite{Klasnja2012-wy}.
Compared with traditional cessation approaches such as self-report diaries, counseling, or periodic clinic visits, wearable technologies offer several important advantages.
First, they enable passive and continuous monitoring, reducing reliance on user memory and self-discipline.
Second, they reduce cognitive burden, an important consideration in real-world cessation contexts where users may already be dealing with stress, work demands, and competing priorities.
Third, they support longitudinal monitoring in naturalistic environments, allowing systems to capture evolving patterns in craving, relapse risk, and intervention effectiveness over time.
These capabilities are particularly important in modern lifestyles where sustained self-regulation is often impractical. By embedding sensing and intervention into daily routines, wearable technologies may improve adherence and enable timely, personalized support~\cite{Dumortier2016-ro, Skinner2019-ad, Cole2017-gl, Senyurek2019-yd, Childs2010-bl}.
This transition toward passive, long-term monitoring also aligns with broader trends in digital health and ubiquitous computing, where wearable systems are increasingly being used to support chronic behavior change.

\subsection{Dataset Diversity and Asian Populations}
A major limitation in existing wearable smoking cessation research is the lack of dataset diversity and geographic representativeness.
Machine learning models trained on narrow demographic groups often generalize poorly to unseen populations, especially when cultural, behavioral, and contextual differences influence smoking pattern~\cite{Fessler2024-ll, Whitney2025-ks}.
Despite the global burden of tobacco use being substantial in Asia, most publicly available smoking-related datasets have been collected in Western populations, particularly in the United States~\cite{Horvath2024-zz, Dumortier2016-ro}.
This creates a significant research gap, as smoking motivations, social norms, work routines, and cessation drivers may differ substantially across Asian populations.
Importantly, smoking prevalence trends across many Asian countries remain stable or increasing, in contrast to the declining trends observed in many Western nations~\cite{drenkow2025detecting}.
Developing datasets and design insights grounded in Asian populations is therefore essential for building robust, equitable, and culturally relevant AI-driven cessation systems.
Our pilot study contributes to this gap by generating early qualitative evidence from a multi-ethnic Asian participant cohort, with a specific focus on culturally grounded craving triggers, social influences, trust, and privacy expectations.

\section{Method}
In this section, we detail the methodology of our study. 
The ethics statement for our pilot study is presented in Appendix~\ref{appendix:ethics}.

\subsection{Participants}

A total of 10 participants were recruited for this pilot study. All participants were male and ranged in age from 19 to 60 years (M = 32.9, SD = 12.6).
The sample comprised participants from diverse ethnic backgrounds, including five Chinese, two Malay, one Caucasian, and two participants of other ethnicities. 
In terms of nationality, the cohort included three Singaporeans, four Indonesians, and three Chinese nationals.
Educational attainment varied across participants: three held a Master’s degree, four held a Bachelor’s degree, and three had completed high school or junior college education.
Monthly income levels were distributed as follows: one participant reported an income between SGD 6,000–10,000, two between SGD 4,000–6,000, and the remaining participants below SGD 4,000.
Regarding smoking status, six participants were active smokers and four were ex-smokers. Ex-smokers are interviewed for their insight on successful smoking cessation. Smoking histories varied substantially, ranging from 4 to 53 years, with average cigarette consumption ranging from 1 to 60 cigarettes per day. Among active smokers, current motivation to quit smoking was measured on a 10-point scale, yielding a mean score of 4.0 (SD = 2.45).
Studies such as [1] have shown that smoking behavior is highly affected by variables that we collected such as ethnic background, residential place, income level, and education level. This diversity in smoking history, socioeconomic background, and ethnicity is desirable to better capture barriers and facilitators to smoking cessation within a multi-ethnic Asian context. 

Participants were recruited through purposive convenience sampling via announcements distributed across multiple online communities using WhatsApp and Telegram.
To ensure diversity in lived experiences and social contexts, recruitment messages were disseminated across a broad range of communities, including:
\begin{itemize}
    \item National University of Singapore (NUS) student groups
    \item Laboratory and research group networks
    \item Indonesian community groups in Singapore
    \item Music communities
    \item Gaming communities
    \item Sports communities
    \item Church communities
    \item Family and personal networks
\end{itemize}
This recruitment strategy was designed to capture a wide range of smoking experiences, social influences, and attitudes toward wearable health technologies.

\subsection{Iterative Interview Design and Procedure}
The interview protocol was developed iteratively to address two research questions:

\textbf{RQ1:} What are the barriers and facilitators to effective smoking cessation in a multi-ethnic Asian population?

\textbf{RQ2:} What are the desired design features of wearable smoking cessation systems that specifically target craving management rather than post-smoking intervention?

To address \textbf{RQ1}, the interview guide was informed by established psychological frameworks related to smoking behavior, craving, relapse, and cessation barriers [2,3]. These frameworks served as a conceptual scaffold for developing interview prompts on triggers, motivation, support systems, and relapse experiences. The list of the frameworks are attached in Appendix 1.
To address \textbf{RQ2}, participants were first invited to discuss their ideal smoking cessation support strategies in an open-ended manner. The discussion then progressively introduced the role of technology, followed by reactions to the proposed smartwatch-based craving intervention concept. A co-design approach was adopted by actively involving participants in feature ideation, trust considerations, personalization needs, and intervention preferences.
The study employed four iterative rounds of semi-structured interviews, with refinements made after each round based on emerging themes and methodological observations.
\section{Pilot Studies}



\subsection{Iteration 1 (Participants ID01–ID02).}
The first iteration involved two participants.
The initial protocol was structured around a predefined 3$\times$3 framework examining biological, psychological, and sociological factors across the phases before, during, and after craving episodes.
However, this structure proved overly rigid and constrained participants’ ability to narrate their lived experiences naturally. Several cells in the framework were found to be irrelevant to individual participants, which limited the depth and authenticity of responses.
As a result, subsequent iterations shifted toward a more narrative-driven interviewing style, allowing participants to tell their stories more freely while enabling targeted probing when themes relevant to the research questions emerged.
A key refinement was changing question phrasing from cognitively framed prompts such as “What do you think…?” to more experience-centered prompts such as “How do you feel about…?” to encourage richer emotional and experiential responses.

\subsection{Iteration 2 (Participants ID03–ID04).}
The second iteration involved two participants.
This round confirmed that many cessation drivers and barriers emerged more naturally through open storytelling than through highly segmented questioning.
A recurring theme identified during this stage was the critical role of social support systems, including spouses, children, romantic partners, and close family members. These support figures served both as motivational drivers for quitting and as ongoing emotional support during quit attempts.
Additionally, three of the first four participants expressed concerns regarding the potential accuracy of smartwatch-based craving detection algorithms. This led to the explicit incorporation of questions related to system trust, perceived reliability, and false detections in later interviews.

\subsection{Iteration 3 (Participants ID07–ID09).}
The third iteration involved three participants.
A strong behavioral pattern emerged linking smoking habits to recurring social and environmental contexts, particularly workplace smoke breaks, social drinking, and peer interactions.
These findings prompted refinements to probe the role of habitual and context-dependent triggers, especially those embedded within daily routines and social rituals.
Participants also highlighted the motivational value of reward-based interventions, particularly for individuals with competitive personalities. This led to the exploration of potential features such as:
small financial incentives or vouchers,
peer leaderboards among users attempting to quit together, or
progress indicators such as lung capacity improvements.

\subsection{Iteration 4 (Participants ID05–ID06, ID10)}
The fourth iteration involved three participants.
This phase revealed several important design considerations.
First, participants emphasized the importance of discussing all quit attempts, including unsuccessful or short-lived ones, as these often provided valuable insights into relapse triggers and opportunities for intervention.
Second, nearly all participants emphasized the need for personalization, both in terms of interface design and intervention strategies aligned with individual motivations and preferences.
Third, privacy concerns became increasingly salient. Participants expressed discomfort with the extent to which the smartwatch might infer their activities, including location-based habits, eating routines, and hand gesture patterns.
Finally, multiple participants emphasized that the system should adopt an empowering, supportive, and non-judgmental tone, rather than appearing condescending or overly patronizing.
These findings directly informed the final interview guide and the design principles for the proposed system.

\subsection{Final Interview Flow}
Following the four iterative rounds, a final semi-structured interview flow was established.
The final protocol was intentionally designed to remain loosely structured yet sufficiently focused, enabling participants to freely describe their lived experiences while ensuring alignment with the study objectives.
The interview consisted of five stages:

\subsubsection{Introduction}
\begin{itemize}
    \item Study overview and purpose
    \item Voluntary participation and informed consent
    \item Opportunity for participant questions
\end{itemize}

\subsubsection{Opening Questions}
\begin{itemize}
    \item Smoking history and personal journey
    \item Previous quit attempts
    \item Role of self, family, friends, and colleagues in smoking behavior
\end{itemize}

\subsubsection{Core Smoking Questions}
\begin{itemize}
    \item Triggers experienced before smoking
    \item Emotions and experiences during a craving episode
    \item Interventions found to be effective
    \item Relapse prevention strategies
    \item Advice from ex-smokers
\end{itemize}

\subsubsection{Smartwatch and Co-Design Questions}
\begin{itemize}
    \item Ideal intervention during a craving
    \item Imagined role of smartwatch and smartphone support
    \item Reactions to craving-detection interventions
    \item Desired features
    \item Trust requirements
    \item Preferred intervention modalities (e.g., mindfulness, rewards, social support)
\end{itemize}

\subsubsection{Closing Questions}
\begin{itemize}
    \item Additional reflections
    \item Changes in perception toward wearable cessation technology
    \item Willingness to use the wearable in daily life
    \item Surprising insights from the interview
\end{itemize}

\begin{figure}[ht]
    \centering
    \includegraphics[width=0.5\linewidth]{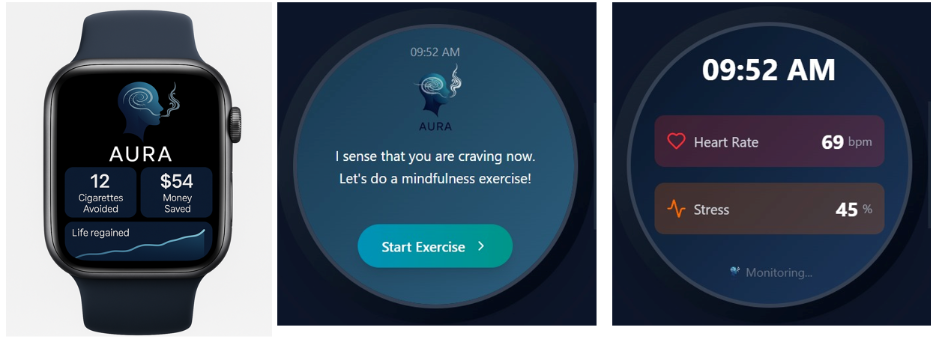}
    \caption{Visual Prototype of AURA}
    \label{fig:AURA_interface}
\end{figure}

Participants were shown a visual prototype of the smartwatch interface (Figure \ref{fig:AURA_interface}) during the technology-focused section to facilitate concrete feedback and co-design discussion.
A full document outlining the standardized participant interview procedure, including privacy protection measures for involved participants, can be found \href{https://drive.google.com/file/d/1K_wxc0j3V9VJdYTHrpHbHPQjIZE8b6HP/view?usp=drive_link}{in this linked PDF}.

\section{Results}
We organize the findings from our pilot studies' qualitative interviews in a thematic analysis over three overarching components of user behavior: predictors of smoking craving, effective interventions for cessation, and design considerations for the usage of health wearables.

\subsection{Predictors of Smoking Craving}
\label{predictors}
Our interviews with participants revealed that smoking is habitual, but also that these habits are specific to the specific individual. Participants 1 summarized this with his insight that smoking “is a habit” which he “cannot stop”, but also that “every people will be different. Their craving situations, their addictions, their thinking, their reason, very very different”. Nonetheless, we found that, in spite of individual differences, there are groups of emotional, social, and contextual habitual triggers that are reliably associated with smoking among our participants.  

\subsubsection{Emotional}

Individuals’ craving to smoke can arise from specific emotions. This means that some smokers tend to associate the act of smoking with specific emotional triggers which, in turn, elicit the craving to smoke. Most participants identified negative emotions as a trigger to smoke. For example, Participant 3 shared that “work stress” triggers a craving to smoke: “I don’t know how to describe it, but it feels a bit better and you don’t feel so agitated. You calm yourself down in a sense and you’re able to think properly”. However, although most participants identified stress as a trigger to smoke, other negative emotions, or even no emotion at all, can also trigger craving. For example, Participant 5 shared that “when it’s like a bit gloomy, like a bit like raining outside, like that would make me really want to smoke. Participant 4 shared that “sometimes, it’s literally when you’re bored”:

\begin{quote}
    “Like when you’re home at night and you’re alone with your thoughts, you know, like you’ve turned off your computer, you’re not doom scrolling. You’re just lying there awake and you’re just thinking about it and you’re like, fuck, a cigarette would be really good right now, you know?”
\end{quote}

\subsubsection{Social}

Individuals’ craving to smoke can arise from specific social situations. This means that some smokers tend to associate the act of smoking with specific social triggers which, in turn, elicit the craving to smoke. For some participants, these social triggers occur during work. Participant 3 shared that “the main reason it’s very difficult” for him to stop smoking is because his colleagues “have informal meetings at the smoking area” which are “sometimes very critical because a lot of information is actually shared during this kind of informal sessions”. For other participants, these social triggers occur outside work. Participant 4 shared that both his girlfriend and him “like social smoke”, while Participant 5 admitted that his friends “definitely do have an impact on [his] smoking habits”, especially since “they were the ones who got [him] started”. Participant 6 shared that “if I go back for reservist or whatever right, I mean there’s always a bit of that temptation [...] even if I don’t bring in right, like people will still offer”. 

\subsubsection{Contextual}

Individuals’ craving to smoke can arise from specific contexts. This means that smokers tend to associate the act of smoking with specific contextual triggers which, in turn, elicit the craving to smoke. Participants 1, 4, 6, 7, 8, 9, and 10 shared similar habits of smoking after activities such as commuting, eating meals, going for toilet breaks, and leaving work. For example, Participants 9 has associated driving long distances with smoking. Participant 4 said, “one cigarette after breakfast, one cigarette after, you know, lunch, and then one when you’re leaving the office”. Participant 1 identified “two hard moments” during which the intensity of smoking is highest: “after eating and during the toilet”. Participant 1 identified “two hard moments” during which the intensity of smoking is highest: “after eating and during the toilet”. Participant 4 used this knowledge to his advantage. Recognizing that his brain had associated the game Dota, he worked toward breaking the link:

\begin{quote}
    “I started playing a different game and I took a long break from Dota. So that's an example. Because my brain associates Dota games or like any mobile game that's half an hour to 45 minutes with smoking. So I had to take a break from Dota. Otherwise, every time I would be at the post-match screen, my brain would automatically think cigarette, cigarette, cigarette.”
\end{quote}

\subsection{Interventions for Smoking Cessation}
\label{interventions}
A number of participants expressed interest in technological-mediated interventions for smoking cessation, emphasising that they can augment, rather than substitute, existing techniques that had helped them in the past. Most notably, Participant 1 remarked how such technologies could have expedited the process of him quitting smoking:

\begin{quote}
    "I think wearing the smartwatch will help me faster the momentum of stop smoking. [...] Like if a long time ago I stopped smoking in one year, maybe when I use my watch, my digital watch that you guys making, I can stop maybe for seven months, eight months. It can reduce the time because it makes me more aware about stop smoking.”
\end{quote}

Reviewing participants’ interviews, we found that participants’ suggestions centered on techniques which they had already adopted in the past, but which they felt could have been made easier with the help of technology. These suggestions focused on techniques on distraction, accountability, achievement tracking, and providing a “voice of reason”.

\subsubsection{Distraction}

Interventions that use distraction divert users’ attention away from smoking to another activity. Participants shared different strategies to divert their attention away from smoking. For example, Participant 1 recognised that meals are a contextual trigger for his craving to smoke, and he video calls his son: “my son can see me and I don’t smoke after eating or during eating. I usually video call with my family, my son. So, yeah, I think it’s very effective.” Participant 1, together with 3 and 4 also reported distracting themselves with push-ups, sit-ups, sweets, or alcohol. On the one hand, Participant 3 shared that “the adrenaline [from push-ups and sit-ups] replaces [nicotine], while Participant 4 recognized that both his dad and him “replaced a desire for the nicotine with a desire for sugar” or “alcohol”: “if there’s nothing wholesome to replace the void, they will replace one addiction just with another. That’s easy”. In order for distraction to work as an effective intervention, Participant 10 suggested that the wearable can provide recommendations, while Participant 1 emphasized that any such recommendation had to be “personalized”: “What will help you if you want to change your smoking into another situation or another activity?” For some participants, such as Participant 10, these recommendations could include “relaxing” activities such as “mindfulness”, “taking a quick nap” or “playing your phone”. 

\subsubsection{Accountability}

Interventions that use accountability share users’ successes and failures with a trusted other. Participant 1 shared his experience of holding himself accountable to his wife, who supported him throughout his smoking cessation journey:

\begin{quote}
    “I only tell my wife because my wife told me, you have to stop smoking, but if you’re still smoking today or tomorrow, you have to tell me, you have to be honest. So I told her because, if I told her, like, for example, today I smoke three times,for example, and tomorrow I only smoke two times, it is a big, what is it, big things for me in front of my wife. So my wife will be happy because, yeah, yesterday you spent three.”
\end{quote}

Other participants, such as Participant 5, shared their experience with “friends [who] wanted to quit like altogether”, although it “lasted probably a week”. Despite participants’ general positive impression of accountability interventions, they also highlighted problems with implementing these techniques on their own. For example, Participant 1 admitted that “nobody can be honest in the first time”, and Participant 10 admitted that “actually I a bit lied to my ex-girlfriend. I don’t smoke in front of her. But after I go home, yeah, of course”. Given these challenges, some participants welcomed the idea of wearables that enforce accountability, especially because it means that they “cannot run” (Participant 3):

\begin{quote}
    “You’d be disappointing the people who supported you. If you do it [i.e., accountability intervention], that’s a very strong way to curb yourself from letting up. That means when you light up, somebody will know. The way you can hide, you know a lot of times you probably can hide in one corner, nobody knows, wash your hands, rinse your mouth, and all that. But this one is like, cannot run. Yeah, it helps. I think it helps.”
\end{quote}

\subsubsection{Achievement Tracking}

Interventions that use achievement tracking help users keep track of important aspects of their lives have improved as a result of quitting smoking. These aspects of their lives include, but are not limited to, “money to buy tobacco product” (Participant 2), improvements in “heart rate”, “blood flow” or “oxygen” (Participant 3), or the increased that they can spend with their loved ones (Participant 1). However, participants highlighted several challenges with implementing these interventions on their own. For example, when Participant 4 attempted to track changes in his physical health, he noted that “you don’t really feel the difference, at least not immediately”. Participant 3 recognised that, for smokers who are more financially comfortable, they can be desensitized to the cost of cigarettes, while Participant 4 admitted that improvements in social relationships, while “noble”, might not actually be the main motivation for quitting. To the extent that these achievements are meaningful to users, participants have welcomed the use of wearables to help them track changes in their physical health (e.g., Participant 3), reminding them of “money saved” as a “smart psychological trick” (e.g., Participant 5), and even “amount of life regained from not smoking” (e.g., Participant 2). Participant 4 suggested that such forms of tracking can be gamified for users who are “driven by that need to want to achieve something”.

\subsubsection{Voice of Reason}

Some users expressed how, in spite of knowing the drawbacks of smoking during moments of clarity, they temporarily forget these harms when craving intensifies. These users expressed a desire for a device that can provide a “voice of reason” during such situations. Participant 3 summarizes how reason often fails without external help, especially during moments of craving:

\begin{quote}
    “There’s always good and bad, pros and cons. But when you’re facing a situation where you’re like craving, right? So you become weak. Your mind becomes weak. So you think of all the good. Like, I have friends here, we can chat, we can talk. Then you forget about all the bad.”
\end{quote}

To this end, he suggested that such wearables can “warn him, show him some some passage about the detrimental effects of tobacco on the health, or just a news article of this guy who died from smoking”. Nonetheless, the use of such scare tactics can backfire on some users. Recognizing that such forms of interventions can sometimes devolve into something overly paternalistic, Participant 4 emphasised the need for the message to be neutral and non-judgmental.

\subsection{Design Considerations for Health Wearables}
\label{designconsid}
Participants proposed a number of design considerations for health wearables which centered on the themes of personalization, privacy, trust, and empowerment. It is important to note that these themes share significant overlap, and that design features can address multiple considerations simultaneously.

\subsubsection{Personalization}

Health wearables should be personalized. This means that wearables should be informed by individual users’ profiles. Participant 4 summarized this principle as follows: “what works for me might not work for the next guy you interview, but if your app or if your smartwatch thing had both of these [options], then I think that covers most of the bases.” For example, although he preferred one that “sound[s] like a caring big brother or like a caring friend as opposed to like a tiger mom”, Participant 3 preferred warnings “about the detrimental effects of tobacco on the health” and even “electric shocks”. Participant 10 warned that wearables should be informed by individual users’ current, and not past, profiles. For example, although enlisting the help of loved ones “could be an innovation”, doing so for users who “don’t have loved ones” or with “ex-girlfriends” can be “very insulting”.

\subsubsection{Privacy}

Health wearables should be private. This means that wearables should respect individual users’ rights to control the collection, use, and disclosure of their personal information. In general, users agreed that wearables should “always [have] that prompt, whether [users] agree to track or all that kind of stuff” (e.g., Participant 6). In addition, even if users agree to track their information, there should also “be a feature that they can turn on and turn off” (Participant 4). Participant 4 compared this to “how a lot of smart watches have like an auto exercise detect feature, where like if you just start walking, it’ll just start logging it as an exercise under walking and you can turn it off”. Participant 10 observed that wearables that enlist the help of loved ones should nonetheless “give suggestion before [they] notify our parents or girlfriend”, while Participant 6 warned that people who “don’t even declare that they smoke” would use these wearables only if “the app icon itself is [not] a giveaway”.

\subsubsection{Trust}

Health wearables should be accurate. This means that the wearables’ outputs (e.g., descriptions, diagnoses, predictions, prescriptions) should correspond to ground truth and be free from systematic errors. A number of participants expressed skepticism that wearables can detect their cravings (e.g., Participants 3, 4, 5, and 10). Participant 4 warned to “implement [predictive] features at [our] own peril” because the wearable “has to be very sure that you were indeed smoking” and that he “would be very pissed off and probably uninstall the app if [he] got a false positive” after a long period of abstinence. Participant 10 was “worried” and “concern[ed]” that “if it’s not accurate, it will be disturbing [him].” Participant 3 shared that “the only way we can trust it is when it’s proven, it’s verified”, and Participant 5 elaborated that such proofs and verifications can come in the form of “endorsements from like the Ministry of Health or something”.

\subsubsection{Empowerment}
\label{designconsid_empowerment}
Health wearables should be empowering. This means that wearables should respect users’ agencies and allow them to actively, instead of passively, shape their smoking journey. Although not all participants were against “tiger mom” wearables (e.g., Participant 3), as described above, most participants agreed that these wearables should augment or support, instead of substitute, users’ roles as active agents in their smoking cessation narrative. Participant 5 shared that the messages should not appear to be “coming from someone with a condescending tone” but “a neutral object telling you that, okay, what you’re doing is not right”. Participant 4 agreed that he “might have a sort of compulsion to want to turn it off because it just starts sounding like a nagging parent”, but he also advised that it is “just a matter of the phrasing of it being like not patronizing”. In addition to customising the “phrasing”, Participant 6 also recommended to “let [users] pick the activity so at least if they give the app or the wearable a chance [...] at least they know the alternative is something that they’re comfortable to do”.

It is noteworthy that not all participants welcomed technological solutions to support their smoking cessation journey, citing reasons such as a preference for “quick fixes” which wearables cannot provide, as well as a general skepticism over the effectiveness of such devices in the absence of users’ own willpower. For example, although Participant 3 was open to using such devices, he shared that “the fastest way is just inject something into me and then you have no more craving, if that’s possible”. Participant 8 felt that because smoking addiction is inherently biological, wearables which fail to intervene at the molecular level by “preemptively occupying the nicotine receptors” are “impossible to effectively relieve nicotine withdrawal symptoms”. Participant 9 remarked that “quitting smoking mainly relies on one’s own willpower”, and that “other products and things are unreliable”. Beyond these reasons, other participants cited reasons such as difficulty in getting people to download apps (Participant 6), or “wearing smartwatches daily” (Participant 5), as potential reasons that might limit adoption.
\section{Discussion}
In this section, we discuss several implications of the qualitative results and resulting themes garnered from the pilot study. 

\subsection{Mapping Predictors to Tailored Interventions}
To maximize the effectiveness of interventions, it may be helpful to tailor intervention types towards specific predictors associated with individual smoking behavior~\cite{nahum-shaniJustinTimeAdaptiveInterventions2016, zhangEffectivenessPersonalizedSmoking2023}.
As the predictor is subject to change over time even within each individual, it becomes vital that intervention types are then adjusted to be delivered at appropriate timings, perhaps given the contextual information that the user may provide.

\textit{Contextual} predictors of smoking may be more effectively targeted by \textit{Distraction}-based interventions. 
According to research on cue-reactivity~\cite{betts2021meta}, the neural pathways associated with these contextual cues can be weakened over time through consistent replacement with unfamiliar behaviors.  
By diverting user attention---or even physical proximity---away from the user's commonly associated cues, we can positively impact user behavior in a passive and seamless manner.

Users experiencing more \textit{Emotional} predictors of smoking may find \textit{Voice of Reason}-based interventions more effective.
CBT~\cite{taylorIntegratingSmokingCessation2025} and Acceptance and Commitment Therapy (ACT)~\cite{mcclureComparativeEffectivenessGroupDelivered2018} have been shown to be effective methods in targeting addictive behavior, with their processes being highly linked to the idea of using verbal motivation to restructure and align one's impulsive, emotional thoughts to that of long-term cessation.
In appropriately employing \textit{Voice of Reason}-based interventions, we may be able to mirror the positive impacts of the established CBT and ACT methods for sustainable behavior change.

Finally, for more \textit{Social} predictors, we may find that \textit{Accountability}-based intervention methods may prove to be more effective.
In cases of users driven by more of peer pressure and the need for belonging, internal reasoning may be less effective than external social accountability systems.
This is supported by SDT~\cite{deci2012self, malinUnderstandingPerspectivesSmoking}, where smokers with a high need for ``relatedness'' thrive more under external accountability models.
With greater need for adherence to external social support, perhaps in the form of family support or group quitting programs, the social urge to smoke may be overtaken by a desire to not let someone else down.
It is also worth mentioning the potential of tethering \textit{Achievement Tracking} interventions to social leaderboards~\cite{chenMobileSocialNetwork2020}, which holds the potential to further anchor these positive maintenance behaviors to a community of like-minded users.
AI agents in these social media settings have been shown to have positive effects~\cite{wangSocialMediaBased2018}, and can be considered to further bolster cessation efforts, particularly in the form of ``companions'' seeking to ensure \textit{Accountability}.

\subsection{Moving from Reactive to Proactive Interventions}
Despite the established benefits of proactive or preventative healthcare solutions~\cite{alqaralehPuffsPredictionsLeveraging2024, vinciProximalEffectsJustinTime2025, maguire2021smoking}, we note that this may not be the definitive path forward in smoking cessation interventions, contrary to our initial hypothesis. 
While a handful of participants explicitly note the general usefulness of proactive solutions in comparison to reactive solutions (P4, P5, P10), several participants also noted risks in doing so. 
P4 mentioned running the risk of a ``forbidden fruit'' situation where one may ``genuinely not have been thinking about it but now you've made them think about it [...] it's like when you tell a kid not to do something---what are they now thinking about? They think about doing it.''
This is aligns with extant research on the prevalence of deviant behavior due to the evocation of excitement and pleasure~\cite{sussmanForbiddenFruitPrediction2010}.
In addition, juvenile smokers may also feel heightened inclinations to smoke despite admonitions about the behavior, owing to the desire to assert signs of independence~\cite{sussmanForbiddenFruitPrediction2010}.
It is thus necessary to consider intricately the user flow involved in the delivery of proactive interventions, to prevent having an effect contrary to what was intended.

An insightful point raised by several participants (P4, P6) was the relevance of including both preemptive and post-hoc intervention methods in a cessation system. 
P4 mentioned ``covering more bases'', a sentiment also echoed by P6 to ``target everyone''. 
In their consideration for targeting a wider potential audience, P6 cited reasons such as individual differences in: level of dependence on smoking, or psychological factors like ``awareness'', ``motivation'' or ``willpower''.
Where preemptive methods may work for those with lower smoking dependence, those with higher dependence may directly overlook them methods due to overpowering urges.
In such cases, it may be more appropriate to have a fail-safe in the form of post-hoc reminders, which can then gradually transition to preemptive methods as the user's reliance dips.
Alternatively, it is possible to consider increasing the severity of preemptive interventions like choosing to employ \textit{Accountability} instead of mere \textit{Distraction}, or even increasing the intensity of \textit{Voice of Reason} prompts.

\subsection{Is There a Need for a Technosolution?}
While we propose a proactive approach with AURA aligning with burgeoning studies on preemptive health interventions~\cite{maguire2021smoking, vinciProximalEffectsJustinTime2025}, a fundamental question must be addressed:
is a technological solution genuinely necessary for smoking cessation?
Our qualitative findings reveal nuanced perspectives on this issue.
Notably, several participants expressed explicit skepticism regarding the efficacy of digital interventions, such as with P9 inclined to push through with ``willpower'', and P8 mentioning the biological limitations of non-intrusive technological solutions. 
Even those who expressed a desire for support from wearables during moments of intense craving had concurrent fears of over-reliance and the erosion of personal autonomy, like P4 warning about psychological reactance~\cite{brehm2012psychological}. 

However, these views need not negate the value of a technological solution; rather, they stand to define the solution's boundaries.
Traditional cessation methods tend to suffer from long-term retention because they rely on sparse, episodic engagement~\cite{Agboola2015-us, Yasin2012-lk} or demand unsustainable cognitive load from the user~\cite{bottorffFathersNarrativesReducing2009}.
Wearable technology could potentially address this gap effectively by seamlessly integrating into routine user behaviors as a catalyst for behavior change, rather than taking center stage in the cessation process.
Aligning with the design considerations proposed in Section \ref{designconsid}, we may be able to translate an abstract, cognitively demanding concept like ``willpower'' into passive awareness empowered by a wearable device.
Empowering the competence and autonomy of users in this manner is vital for sustainable behavioral change, according to SDT~\cite{deci2012self}.
To effectively support users without usurping their agency, technological cessation systems should thus prioritize gentle, actionable insights \textit{personalized} to the user, to reinforce their competence for long-term cessation benefits.

Furthermore, as digital interventions become increasingly widespread and sophisticated, it is crucial to address the potential digital divide inherent in wearable health technologies. 
As seen in our qualitative interviews, not all prospective users possess high levels of technological literacy or innate comfort with smart devices.
AURA should be deliberately designed to accommodate even individuals who may not be open to technological health interventions, otherwise risking the exacerbation of existing health disparities.
A promising design strategy is to ground the digital experience in familiar, analog behavioral methods, as in the ones identified in Section \ref{interventions}.
By translating established, real-world coping strategies like \textit{Accountability} to a spouse or engaging in physical \textit{Distractions} into that of digital interactions, we may be able to reduce the cognitive load and learning curve for inexperienced users. 
Ensuring the interface remains accessible and firmly anchored in established psychological practices will be critical for sustained engagement and effective cessation with AURA.


\section{Conclusion}
Smoking cessation remains a complex behavioral challenge, where spontaneous cue-triggered cravings frequently overpower long-term intentions.
Our pilot study illuminates the subtle nuances in addiction management, revealing clusters across emotional, social and contextual domains, as well as various commonly used intervention types.
Crucially, participants emphasized that proactive health wearables should be deeply personalized, protective of user privacy, highly accurate, and empowering rather than dictating user agency.

This qualitative pilot study represents a first step toward ubiquitous, user-centered cessation management systems. 
We intend on experimenting with different configurations of AURA---incorporating various themes identified in the current study---in a high-fidelity setting on a wide range of users, including even those not inclined towards technological solutions, to fully assess its capabilities.
In addition, a critical limitation of our qualitative study is the lack of female users, due to difficulties in recruitment and a significantly higher prevalence of male smokers.
We intend on employing snowball sampling in future studies, which is often used to reach under-represented and hard-to-reach samples. 

\begin{acks}
We thank Prof. Jane and the TAs, Yize and Peinuan, for their guidance throughout this lovely module :)  
\end{acks}

\bibliographystyle{unsrt}
\bibliography{references}

\newpage 
\appendix
\section{Appendix A: Theoretical Frameworks Underpinning Survey Design}
\label{appendix:frameworks}

Our questions and framework are inspired by \cite{luBarriersFacilitatorsSmoking2025, nguyenDesiresNeedsQuitting2024, leppinExploringPerspectivesDigital2025}.
\appendix

\subsection{Perceived Risk and Benefit: Health Belief Model}
\textit{Angle: user belief}

\paragraph{Drivers}
\begin{itemize}
    \item Underestimated risk severity (e.g., ``I will just cough occasionally, I won't get cancer or stroke'')
    \item High perceived benefits (social bonding, stress relief)
\end{itemize}

\paragraph{Inhibitors}
\begin{itemize}
    \item Strong belief in the benefits of quitting
    \item Perceived high risk of continued smoking
    \item External cues such as health warnings, physical symptoms, phone notifications, or social interaction
\end{itemize}

\subsection{Real Reward and Punishment: Operant Conditioning}
\textit{Angle: addiction and habits}

\paragraph{Drivers}
\begin{itemize}
    \item Positive reinforcement (pleasure derived from nicotine)
    \item Negative reinforcement (relief from stress or withdrawal; a quick escape from reality)
\end{itemize}

\paragraph{Inhibitors}
\begin{itemize}
    \item Competing rewards (e.g., receiving a voucher each time a craving is resisted)
    \item Removal of the smoking reward
    \item Introduction of costs (price increases, smoking restrictions)
\end{itemize}

\subsection{Cues: Cue Reactivity / Classical Conditioning}
\textit{Angle: craving episodes}

\paragraph{Drivers}
\begin{itemize}
    \item Environmental triggers: meals, stressful meetings, social settings, caffeine consumption
    \item Cue-induced craving (internal feelings): a sense that something is missing after a meal, anxiety following a stressful meeting, restlessness after witnessing someone smoke
\end{itemize}

\paragraph{Inhibitors}
\begin{itemize}
    \item Avoiding smoking-associated cues
    \item Breaking conditioned associations (e.g., dissociating meals from craving, and craving from smoking) — a core target of our proposed solution
\end{itemize}

\subsection{Learning and Role Models: Social Cognitive Theory}
\textit{Angle: role models and coaching}

\paragraph{Drivers}
\begin{itemize}
    \item Uptake of smoking behavior learned from peers or family members
\end{itemize}

\paragraph{Inhibitors}
\begin{itemize}
    \item Exposure to role models who have successfully quit smoking
\end{itemize}

\subsection{Intrinsic vs.\ Extrinsic Motivation: Self-Determination Theory}
\textit{Angle: long-term behavior change, motivation source and quality}

\paragraph{Drivers}
\begin{itemize}
    \item External pressure to quit (e.g., forced cessation) leading to weak adherence
    \item Smoking as a means of fulfilling psychological needs (stress coping, sense of autonomy)
\end{itemize}

\paragraph{Inhibitors}
\begin{itemize}
    \item Intrinsic motivation to quit (e.g., personal values, health, faith)
    \item Interventions that support autonomy and build competence
\end{itemize}

\subsection{Impulse vs.\ Control: Dual Process Theory}
\textit{Angle: intentions fail due to fast, automated responses}

\paragraph{Drivers}
\begin{itemize}
    \item Automatic impulses (cravings, habit loops) that bypass deliberate decision-making
\end{itemize}

\paragraph{Inhibitors}
\begin{itemize}
    \item Reflective control strategies (planning, self-awareness)
    \item Disrupting automatic responses before they trigger smoking behavior
\end{itemize}

\subsection{Coping: Stress and Coping Theory}
\textit{Angle: coping strategy and relapse factors}

\paragraph{Drivers}
\begin{itemize}
    \item Use of nicotine as an emotion regulation strategy
    \item Stress, anxiety, and negative affect as relapse triggers
\end{itemize}

\paragraph{Inhibitors}
\begin{itemize}
    \item Alternative coping strategies (mindfulness, physical exercise, contacting a caretaker or support person)
    \item Stress reduction interventions
\end{itemize}

\section{Appendix B: Ethics Statement} 
\label{appendix:ethics}

\subsection{Risk 1: Bias and health inequities from non-representative data}
\subsubsection{Risk} 
AURA’s prediction model will be trained on a small pilot dataset drawn from the authors’ social networks in Asia. 
This convenience sample likely skews toward a narrow demographic slice, so the model may work for people like our participants but fail for smokers of different ages, income levels, or cultural backgrounds. 
In practice, some users would get inaccurate predictions: false alarms that cause frustration, or missed cravings when they matter most. 
These errors would fall hardest on groups already least served by existing cessation tools. 
There is also a subtler issue. 
A system that frames smoking as something to be predicted and preempted risks reinforcing the stigma that smokers simply lack willpower. 
In many Asian cultures, smoking is tied to social rituals, masculinity and class, and a poorly calibrated intervention could alienate the very users it aims to help.

\subsubsection{Mitigation}
In future stages we plan to recruit from a wider range of Asian subpopulations and a more representative gender makeup. 
We will report the demographic makeup and known limitations of our pilot data so readers can judge generalizability for themselves, and we will break down model accuracy by demographic group rather than reporting only aggregate numbers. 
After the pilot, we will run feedback sessions with participants to check whether our assumptions actually match their experience. 
AURA’s messaging will draw on Acceptance and Commitment Therapy (ACT), which treats cravings as ordinary experiences rather than personal failings, to avoid reinforcing shame.

\subsection{Risk 2: Privacy erosion and misuse of continuous behavioral data}
\subsubsection{Risk}
AURA collects IMU, heart rate, GPS, and smartphone activity data continuously. 
If always-on health sensing becomes normalized, the same approach could be adopted by employers, insurers, or governments to gate access to jobs, coverage, or services based on monitored behavior. 
Lower-income smokers, who have less leverage to refuse such monitoring, would be most exposed. 
The prediction methodology itself is also dual-use: a commercial actor could time advertisements to predicted craving moments, or the pipeline could be adapted to exploit other urges like gambling or compulsive spending.

\subsubsection{Mitigation}
Pilot study sensor data will be stored on encrypted, access-controlled university servers, with GPS coordinates and timestamps anonymized before analysis. 
We will not release raw data publicly; shared datasets will be aggregated and de-identified, and model access will require a data use agreement limited to research purposes. 
Users will always be able to disable monitoring or dismiss prompts, and all interventions will be framed as suggestions. 
Our paper will discuss dual-use risks and explicitly state that craving prediction data should not be repurposed beyond personal cessation goals.

\subsection{Risk 3: Over-reliance and automation bias}
\subsubsection{Risk}
If users trust AURA’s predictions too much, they may stop building their own coping strategies. 
A missed prediction could leave someone unprepared at exactly the wrong moment, worsening a relapse. 
If the system goes offline, a user who has outsourced craving management to the app has no fallback. 
More broadly, convenient AI health tools risk becoming substitutes for professional cessation support like counseling or medication, when they should be complements.

\subsubsection{Mitigation}
AURA’s interventions will teach coping techniques, not just deliver alerts, so that users get better at managing cravings over time rather than becoming more dependent on the system. 
Predictions will show confidence levels to make clear they are probabilistic, not certain, and the app will state plainly that it does not replace professional medical care. 
Our writeup will cover automation bias as a known risk in health AI and recommend that future deployments periodically prompt users to check in with a healthcare provider.

\end{document}